\documentclass[11pt]{article}
\usepackage{graphicx}
\usepackage{amsfonts}
  \def\fX{{\cal X}}

\usepackage{theorem}
\theorembodyfont{\rmfamily}
\newtheorem{Lem}{Lemma}[section]
\newtheorem{Def}[Lem]{Definition}
\newtheorem{The}[Lem]{Theorem}
\newtheorem{Prop}[Lem]{Proposition}
\newtheorem{Cor}[Lem]{Corollary}

\newtheorem{Rem}[Lem]{Remark}

\newcommand{\qed}{\hbox{\rule{6pt}{6pt}}}

\begin{document}
\title{Information theoretical properties of Tsallis entropies}
\author{Shigeru FURUICHI$^1$\footnote{E-mail:furuichi@ed.yama.tus.ac.jp}\\
$^1${\small Department of Electronics and Computer Science,}\\{\small Tokyo University of Science, Yamaguchi, 756-0884, Japan}}
\date{}
\maketitle
{\bf Abstract.} 
A chain rule and a subadditivity for the entropy of type $\beta$, 
which is one of the nonadditive entropies, were derived by Z.Dar\'oczy. 
In this paper, we study the further relations among Tsallis type entropies which are typical nonadditive entropies. 
The chain rule is generalized by showing it for Tsallis relative entropy and the nonadditive entropy.
We show some inequalities related to Tsallis entropies, especially the strong subadditivity for Tsallis type entropies and 
the subadditivity for the nonadditive entropies. 
The subadditivity and the strong subadditivity naturally lead to define 
Tsallis mutual entropy and Tsallis conditional mutual entropy, respectively, and then we show again chain rules for Tsallis mutual entropies. 
We give properties of entropic distances in terms of Tsallis entropies.
Finally we show parametrically extended results based on information theory.

\vspace{3mm}

\vspace{3mm}

\vspace{3mm}


\section{Introduction}
For the probability distribution $p(x) \equiv p(X=x)$ of the random variable $X$, Tsallis entropy was defined in \cite{Tsa} : 
\begin{equation}\label{def_Tsallis}
S_q(X) \equiv -\sum_{x} p(x)^q \ln_q p(x),\quad (q\neq 1)
\end{equation}
with one parameter $q$ as an extension of Shannon entropy, where $q$-logarithm function is defined by $\ln_q(x) \equiv \frac{x^{1-q}-1}{1-q}$
for any nonnegative real number $x$ and $q$. 
Tsallis entropy plays an essential role in nonextensive statistics so that
many important results have been published in mainly statistical physics \cite{AO}. 
In particular, see \cite{TGS,Tsa1} for the recent results on Tsallis entropies.
The field of the study on Tsallis entropy and related fields 
in statistical physics are often called Tsallis statistics or nonextensive statistical physics due to the nonextensivity Eq.(\ref{pseudo}) in the below.
We easily find that $\lim_{q \to 1} S_q(X) = S_1(X) \equiv -\sum_{x} p(x) \log p(x)$, since $q$-logarithm converges 
to natural logarithm as $q \to 1$. After the celebrated discover by Tsallis, it was shown that
Tsallis entropy can be uniquely formulated by the generalized Shannon-Khinchin's axioms in \cite{San,Ab1,Su3}. 
The axiomatical characterization of Tsallis entropy was improved and that of Tsallis relative entropy was formulated in \cite{Fur}.
It is a crucial difference of the fundamental property between Shannon entropy and Tsallis entropy that
for two independent random variables $X$ and $Y$, Shannon entropy $S_1(X)$ has an additivity : 
\begin{equation} \label{additive}
S_1(X\times Y) = S_1(X) + S_1(Y),
\end{equation} 
however, Tsallis entropy $S_q(X), (q\neq 1)$ has a pseudoadditivity :
\begin{equation} \label{pseudo}
S_q(X\times Y) = S_q(X)+S_q(Y) +(1-q)S_q(X)S_q(Y),
\end{equation}
where $S_q(X\times Y)$ is Tsallis joint entropy, which will be defined in the below, 
for the direct product $X \times Y$ of two independent random variables $X$ and $Y$. 
The independence between $X$ and $Y$ phyisically means that there is no interaction between two systems $X$ and $Y$.
This pseudoadditivity of Tsallis entropy in Eq.(\ref{pseudo}) originally comes from that of $q$-logarithm function $\ln_q(x)$.
We pay attention whether an additivity (such as Eq.(\ref{additive})) of entropies for two independent random variables holds or not (such as Eq.(\ref{pseudo})).
An entropy is called a nonadditive entropy if it does not satisfy the additivity such as Eq.(\ref{additive}), 
while an entropy is called an additive
entropy if it satisfies the additivity specified in Eq.(\ref{additive}). 
The entropy of type $\beta$ introduced by Dar\'oczy in \cite{Dar} (also called the structural $a$-entropy introduced by Havrda and Charv\'at in \cite{HC}):
$$
H_{\beta} (X) \equiv \frac{\sum_{x}\left(p(x)^{\beta}-p(x)\right)}{2^{1-\beta}-1}, \quad (\beta >0, \beta \neq 1)
$$
is a nonadditive entropy. 

We are now interested in Tsallis entropy which is a typical nonadditive entropy.
From Eq.(\ref{pseudo}), we easily find that we have a subadditivity of Tsallis entropies for two independent random variables $X$ and $Y$ in the case
of $q \geq 1$ since Tsallis entropy is nonnegative. 
In the first of the section 3, we will review a subadditivity of Tsallis entropies for general random variables $X$ and $Y$ 
in the case of $q \geq 1$. 

From the entropy theoretical point of view, for the related entropies such as Tsallis conditional entropy and Tsallis joint entropy 
were introduced in \cite{BRT,AR,Ya} and then some interesting relations between them were derived. In this paper, in order to
derive the relation between them, we adopt the following definition of Tsallis conditional entropy and Tsallis joint entropy.
\begin{Def}
For the conditional probability $p(x\vert y) \equiv p(X=x \vert Y=y)$ and the joint probability $p(x,y) \equiv p(X=x,Y=y)$, 
we define Tsallis conditional entropy and Tsallis joint entropy by
\begin{equation}  
S_q(X\vert Y) \equiv  -\sum_{x,y} p(x,y)^q \ln_q p(x\vert y),  \quad (q \neq 1), \label{jc} 
\end{equation}
and
\begin{equation}
S_q(X,Y) \equiv  -\sum_{x,y} p(x,y)^q \ln_q p(x,y), \quad (q \neq 1).  \label{jc2} 
\end{equation}
\end{Def}
We note that the above definitions were essentially introduced in \cite{Dar} by
\begin{eqnarray*}\label{Dar_joint_cond}
&& H_{\beta}(X, Y) \equiv  \frac{\sum_{x,y}\left(p(x,y)^{\beta}-p(x,y)\right)}{2^{1-\beta}-1}, \quad (\beta >0, \beta \neq 1) \\
&& H_{\beta}(X\vert Y) \equiv \sum_y p(y)^{\beta} H_{\beta} (X\vert y),  \quad (\beta >0, \beta \neq 1)
\end{eqnarray*}
except for the difference of the multiplicative function. And then a chain rule and a subadditivity:
\begin{eqnarray*}
&& H_{\beta}(X,Y) = H_{\beta}(X) + H_{\beta}(Y\vert X),\\
&& H_{\beta}(Y\vert X) \leq H_{\beta}(Y), \quad \beta >1,
\end{eqnarray*}
 were shown in Theorem 8 of \cite{Dar}.
In this paper, we call the entropy defined by Eq.(\ref{def_Tsallis}), Eq.(\ref{jc}) or Eq.(\ref{jc2}) 
Tsallis type entropy.
It is remarkable that Eq.(\ref{jc}) can be easily deformed by
\begin{equation} \label{our_con}
S_q(X\vert Y) = \sum_y p(y)^q S_q(X\vert y).
\end{equation}
Note that for $q \neq 1$ we have $S_q(X\vert Y) = \sum_y p(y)^q S_q(X) \neq S_q(X)$ when $X$ and $Y$ are indepentent each other (i.e., $p(x\vert y) = p(x)$ for all $x$ and $y$).
Eq.(\ref{our_con}) can be further regarded that the following expectation value :
$$
E_q(X) \equiv \sum_i p_i^q x_i
$$
 depending on the parameter $q$ is adopted.
Since the expectation value $E_q(X)$ has a crucial lack that it dose not hold $E_{q}(1) = 1$ for $q \neq 1$ in general,
a slightly modified definitions, the normalized Tsallis entropies such as 
\begin{equation}
S_q^{nor}(X) \equiv \frac{-\sum_x p(x)^q \ln_q p(x)}{\sum_x p(x)^q}, \quad (q \neq 1)
\end{equation}
are sometimes used \cite{AR,Ya}, and then important and meaningful results in physics are derived for the normalized nonadditive conditional entropy defined in \cite{Ab1,AR}. 
However, as pointed out in \cite{Abe2}, the normalized Tsallis entropies are not stable, but the original Tsallis entropies are stable as 
experimental quantities so that it is meaningful to study the original Tsallis entropies from the physical point of view. 
Inspired by these pioneering works \cite{Ab1,AR,Abe2}, we therefore study the information theoretical properties of the original Tsallis type entropies without normalization in this paper. 
As advantages in information theoretical study, our approach shows chain rules and subadditivities of the Tsallis entropies
and then naturally leads to define the Tsallis mutual entropy. 
Finally the rough meaning of the parameter $q$ of the Tsallis entropy is given from the information theoretical viewpoint.
Both entropies defined in Eq.(\ref{jc}) and Eq.(\ref{jc2}) recover the usual Shannon entropies as $q \to 1$. 

In the past, some parametrically extended entropies were introduced in the mathematical point of view. Especially, the R\'enyi entropy \cite{R61,R62}
(see also \cite{AD})
\begin{equation}
R_q \left( X \right) = \frac{1}{{1 - q}}\log \sum\limits_x {p\left( x \right)^q } ,\quad (q\neq 1).
\end{equation}
is famous.
For Tsallis entropy and R\'enyi entropy, we have the following relation,
\begin{equation} \label{trans}
R_q \left( X \right) = \frac{1}{{1 - q}}\log \left\{ {1 + \left( {1 - q} \right)S_q \left( X \right)} \right\},\quad (q\neq 1).
\end{equation}
It is notable that R\'enyi entropy has an additivity:
\begin{equation}
R_q(X\times Y) = R_q(X) + R_q(Y),
\end{equation}
for two independent random variables $X$ and $Y$.
Although there exsits a simple transformation between Tsallis entropy and R\'enyi entropy, they have a different structure in regards to the matter of their additivities.
 
This paper is organized as follows. In section \ref{sec2}, we review a chain rule given in \cite{Dar} and give its generalized results. We also give the 
chain rule for Tsallis relative entropy.
 In section \ref{sec3}, we review a subadditivity given in \cite{Dar} and give its generalized results. In addition, we show several inequalities related
Tsallis entropies. In section \ref{sec4}, we introduce Tsallis mutual entropy and then give its fundamental properties. In section \ref{sec5},
we discuss on entropic distances and correlation coefficients due to Tsallis entropies. In section \ref{sec6}, we give some applications of Tsallis entropies
in the context of information theory. Related remarks and discussions are given in section \ref{sec7}.


\section{Chain rules for Tsallis entropies}  \label{sec2}
It is important to study so-called a chain rule which gives the relation between a conditional entropy and a joint
entropy in not only information theory \cite{Cov} but also statistical physics. 
For these Tsallis entropies, the following chain rule holds as similar as 
the chain rule holds for the joint entropy of type $\beta$ and the conditional entropy of type $\beta$. 

\begin{Prop} {\bf (\cite{Dar})} \label{prop1} 
\begin{equation} \label{chain1}
S_q \left( {X,Y} \right) = S_q \left( X \right) + S_q \left( {Y\left| X \right.} \right).
\end{equation}
(Therefore immediately $S_q(X) \leq S_q(X,Y).$)
\end{Prop}
{\bf (Proof)}
From the formulas $\ln_q(xy) = \ln_q x +x^{1-q}\ln_q y$ and $\ln_q\frac{1}{x} = -x^{q-1} \ln_q x$, we have
\begin{eqnarray*}
 S_q \left( {X,Y} \right) &=& \sum\limits_{x} {\sum\limits_{y} {p\left( {x ,y } \right)\ln _q \frac{1}{{p\left( {x ,y } \right)}}} } 
 = \sum\limits_{x} {\sum\limits_{y} {p\left( {x ,y } \right)\ln _q \frac{1}{{p\left( {x } \right)p\left( {y \left| {x } \right.} \right)}}} }  \\ 
 &=& \sum\limits_{x} {p\left( {x } \right)\ln _q \frac{1}{{p\left( {x } \right)}} 
+ \sum\limits_{x} {\sum\limits_{y} {p\left( {x ,y } \right)p\left( {x } \right)^{q - 1} \ln _q \frac{1}{{p\left( {y \left| {x } \right.} \right)}}} } }  \\ 
 &=& - \sum\limits_{x} p\left( x \right)^q \ln _q  p\left( x \right) 
- \sum\limits_{x} p\left( x  \right)^q \sum\limits_{y} p\left( y \vert x \right)^q \ln _q p\left( y \vert x   \right) \\
&=& S_q \left( X \right) + S_q \left( {Y\left| X \right.} \right) . 
 \end{eqnarray*}
\hfill \qed

From the process of the proof of Proposition \ref{prop1},
we can easily find that if $X$ and $Y$ are independent (i.e., $p(y\vert x) = p(y)$ for all $x$ and $y$), 
then Eq.(\ref{chain1}) recovers the pseudoadditivity Eq.(\ref{pseudo}):
\begin{eqnarray*}
 S_q \left( {X,Y} \right) &=& -\sum\limits_{x} {p\left( {x } \right)^q \ln _q {p\left( {x } \right)} 
-\sum\limits_{x} {p\left( {x } \right)^q \sum\limits_{y} {p\left( {y } \right)^q
\ln _q {p\left( {y } \right)}}} }   \\ 
&=& S_q \left( X \right) + \sum\limits_{x} {p\left( {x } \right)^q } S_q \left( Y \right) \\ 
&=& S_q \left( X \right) + S_q \left( Y \right) + \left( {1 - q} \right)S_q \left( X \right)S_q \left( Y \right). 
 \end{eqnarray*}

As a corollary of the above Proposition \ref{prop1}, we have the following lemma.

\begin{Lem}\label{lem2}
The following chain rules hold.
\begin{itemize}
\item[(1)] $S_q(X,Y,Z) = S_q(X,Y\vert Z) +S_q(Z)$.
\item[(2)] $S_q(X,Y\vert Z) = S_q(X\vert Z) +S_q(Y\vert X,Z)$.
\end{itemize}
\end{Lem}
{\bf (Proof)}
By using $\ln_q \frac{y}{x} = \ln_q y+y^{1-q} \ln_q \frac{1}{x}$, we have
\[
S_q \left( {X,Y\left| Z \right.} \right)\, = S_q \left( {X,Y,Z} \right) - S_q \left( Z \right).
\]
In the similar way, we have
\[
S_q \left( {Y\left| {X,Z} \right.} \right) = S_q \left( {X,Y,Z} \right) - S_q \left( {X,Z} \right).
\]
Also from the Proposition \ref{prop1}, we have
\[
S_q \left( {X,Y} \right) = S_q \left( X \right) + S_q \left( {Y\left| X \right.} \right).
\]
Therefore we have

\begin{eqnarray*}
 S_q \left( {Y\left| {X,Z} \right.} \right) &=& S_q \left( {X,Y,Z} \right) - S_q \left( {X,Z} \right) \\ 
 &=& S_q \left( {X,Y\left| Z \right.} \right) + S_q \left( Z \right) - \left\{ {S_q \left( {X\left| Z \right.} \right) + S_q \left( Z \right)} \right\} \\ 
 &=& S_q \left( {X,Y\left| Z \right.} \right) - S_q \left( {X\left| Z \right.} \right). 
\end{eqnarray*}

\hfill \qed

From (2) of of Lemma \ref{lem2}, we have 
\begin{equation} \label{eq_lem22_result}
S_q(X\vert Z) \leq S_q(X,Y\vert Z) 
\end{equation} 

\begin{Rem}
(2) of Lemma \ref{lem2} can be generalized in the following formula :
\begin{equation} \label{eq_remark23}
S_q \left( {X_1 , \cdots ,X_n \left| Y \right.} \right) = \sum\limits_{i = 1}^n {S_q \left( {X_i \left| {X_{i - 1} , \cdots ,X_1 ,Y} \right.} \right)}.
\end{equation}
\end{Rem}

\begin{The} \label{chain_entropy}
Let $X_1,X_2,\cdots ,X_n$ be the random variables obeying to the probability distribution $p(x_1,x_2,\cdots ,x_n)$. Then we have the following
chain rule.
\begin{equation} \label{chain_gen}
S_q \left( {X_1 ,X_2 , \cdots ,X_n } \right) = \sum\limits_{i = 1}^n {S_q \left( {X_i \left| {X_{i - 1} , \cdots ,X_1 } \right.} \right)}. 
\end{equation}
\end{The}
{\bf (Proof)}
We prove the theorem by induction on $n$.
From Proposition \ref{prop1}, we have
\[
S_q \left( {X_1,X_2} \right) = S_q \left( X_1 \right) + S_q \left( {X_2\left| X_1 \right.} \right).
\]
Assuming that Eq.(\ref{chain_gen}) holds for some $n$, from Eq.(\ref{chain1}) we have
\begin{eqnarray*}
 S_q \left( {X_1 ,\cdots  ,X_{n+1} } \right) &=& S_q \left( {X_1, \cdots, X_n } \right) + S_q \left( X_{n+1} \vert X_1, \cdots, X_n  \right) \\ 
 &=& \sum\limits_{i = 1}^n {S_q \left( {X_i \left| {X_{i - 1} , \cdots ,X_1 } \right.} \right)} + S_q \left( X_{n+1} \vert X_n, \cdots, X_1  \right), 
\end{eqnarray*}
which means Eq.(\ref{chain_gen}) holds for $n+1$. 
\hfill \qed
\vspace*{2mm}

In the rest of this section, we discuss on the chain rule for Tsallis relative entropy.
To do so, we use Tsallis relative entropy
\begin{equation}  \label{Tsallisrelativeentropy}
D_q(U\vert V) \equiv - \sum_x u(x) \ln_q \frac{v(x)}{u(x)},
\end{equation}
defined for two probability distributions $u(x)$ and $v(x)$, and $q \geq 0$.
See \cite{Tsa2,Shi,RA,FYK} for details of Tsallis relative entropy.
Eq.(\ref{Tsallisrelativeentropy}) is equivalent to the following forms,
\begin{eqnarray}
D_q(U\vert V) &=& \frac{\sum_{x}\left(u(x)- u(x)^q v(x)^{1-q}\right) }{1-q} \label{non-view-Tsallisrelativeentropy}  \\
&=& \sum_{x} u(x)^q v(x)^{1-q}\ln_q\frac{u(x)}{v(x)}. \label{view-Tsallisrelativeentropy}
\end{eqnarray}

Along the line of the definition of Tsallis conditional entropy Eq.(\ref{jc}), we naturally define Tsallis conditional relative entropy such as
\begin{eqnarray}
D_q \left( u\left( y\vert x  \right) \vert v \left( y\vert x  \right) \right) &\equiv&  \sum_{x,y} u(x,y)^q v(x,y)^{1-q} \ln_q \frac{u(y\vert x)}{v(y\vert x)} \label{def-T-con-rel_ent}\\
&=& \sum_x u(x)^q v(x)^{1-q} \sum_y u(y\vert x)^q v(y\vert x)^{1-q} \ln_q \frac{u(y\vert x)}{v(y\vert x)}
\end{eqnarray}
for two joint probability distributions $u(x,y)$ and $v(x,y)$, and two conditional probability distributions $u(y\vert x)$ and $v(y\vert x)$,
based on the view of Eq.(\ref{view-Tsallisrelativeentropy}), neither Eq.(\ref{Tsallisrelativeentropy}) nor Eq.(\ref{non-view-Tsallisrelativeentropy}).
Taking the limit as $q \to 1$, Eq.(\ref{def-T-con-rel_ent}) recovers the usual conditional relative entropy.
Then we have the following chain rule of Tsallis relative entropy for general (possibly {\it not} independent) case.

\begin{The} {\bf (Chain rule for Tsallis relative entropy)} \label{chainrule-T-relativeentropy}
\begin{equation}\label{chain_rela}
D_q \left( u\left( x,y \right) \vert v\left( x,y \right)  \right) = 
D_q \left( u\left( x \right) \vert v\left( x \right)  \right) + D_q \left( u\left( y\vert x  \right) \vert v \left( y\vert x  \right) \right) .
\end{equation}
\end{The}
{\bf (Proof)}
The proof follows by the direct calculations.
\begin{eqnarray*}
 D_q \left( u\left( {x,y} \right)\vert v\left( {x,y} \right) \right) &=&  - \sum\limits_{x,y} {u\left( {x,y} \right)} \ln _q \frac{{v\left( {x,y} \right)}}{{u\left( {x,y} \right)}}\, =  - \sum\limits_{x,y} {u\left( {x,y} \right)} \ln _q \frac{{v\left( x \right)v\left( {y\left| x \right.} \right)}}{{u\left( x \right)u\left( {y\left| x \right.} \right)}} \\ 
 &=&  - \sum\limits_{x,y} {u\left( {x,y} \right)} \left\{ {\ln _q \frac{{v\left( x \right)}}{{u\left( x \right)}} + \ln _q \frac{{v\left( {y\left| x \right.} \right)}}{{u\left( {y\left| x \right.} \right)}} + \left( {1 - q} \right)\ln _q \frac{{v\left( x \right)}}{{u\left( x \right)}}\ln _q \frac{{v\left( {y\left| x \right.} \right)}}{{u\left( {y\left| x \right.} \right)}}} \right\} \\ 
  &=&  - \sum\limits_{x,y} {u\left( {x,y} \right)} \left\{ {\ln _q \frac{{v\left( x \right)}}{{u\left( x \right)}} + \left\{ {\frac{{u\left( x \right)}}{{v\left( x \right)}}} \right\}^{q - 1} \ln _q \frac{{v\left( {y\left| x \right.} \right)}}{{u\left( {y\left| x \right.} \right)}}} \right\} \\ 
 &=& D_q \left( u\left( x \right)\vert v\left( x \right) \right) - \sum\limits_{x,y} {\left\{ {\frac{{u\left( x \right)}}{{v\left( x \right)}}} \right\}^{q - 1} u\left( {x,y} \right)\ln _q \frac{{v\left( {y\left| x \right.} \right)}}{{u\left( {y\left| x \right.} \right)}}}.  \\ 
&=&  D_q \left( u\left( x \right)\vert v\left( x \right) \right) + D_q \left( u\left( y\vert x  \right) \vert v \left( y\vert x  \right) \right) .
\end{eqnarray*}

\hfill \qed

Taking the limit as $q \to 1$, Theorem \ref{chainrule-T-relativeentropy} recovers Theorem 2.5.3 in \cite{Cov}.
It is known that for $u\left( {y\left| x \right.} \right) = u\left( y \right)$ and $v\left( {y\left| x \right.} \right) = v\left( y \right)$,
Tsallis relative entropy has a pseudoadditivity :
\begin{equation}\label{pseudo_rela}
D_q \left( u(x)u(y) \vert v(x)v(y)   \right) = D_q \left( u(x) \vert v(x)  \right) + D_q \left( u(y) \vert v(y)  \right) 
+ \left( q - 1 \right)D_q \left( u(x) \vert v(x)   \right)D_q \left( u(y) \vert v(y)  \right).
\end{equation}
From the process of the above proof, we find that the chain rule Eq.(\ref{chain_rela}) recovers 
the pseudoadditivity Eq.(\ref{pseudo_rela}) when $u\left( {y\left| x \right.} \right) = u\left( y \right)$ and $v\left( {y\left| x \right.} \right) = v\left( y \right)$.


\section{Subadditivities for Tsallis entropies}  \label{sec3}
From Eq.(\ref{pseudo}), for $q \geq 1$ and two independent random variables $X$ and $Y$, the subadditivity holds:
$$S_q(X \times Y) \leq S_q(X) + S_q(Y).$$
It is known that the subadditivity for general random variables $X$ and $Y$ holds in the case of $q \geq 1$, thanks to the following proposition.

\begin{Prop} {\bf (\cite{Dar})} \label{sub_Dar}
The following inequality holds for two random variables $X$ and $Y$, and $q \geq 1$,
\begin{equation}\label{eq_sub_Dar}
 S_q(X\vert Y) \leq S_q(X),
\end{equation}
with equality if and only if $q=1$ and $p(x\vert y)= p(x)$ for all $x$ and $y$.
\end{Prop}
{\bf (Proof)}
We give the similar proof given in \cite{Dar} as a version of Tsallis type entropies for the convenience of the reader.
We define the function $\hbox{Ln}_q(x) \equiv \frac{x^q-x}{1-q}$ for $q \geq 0, q \neq 1$ and $x \geq 0$. 
Then the function $\hbox{Ln}_q$ is nonnegative for $0 \leq x \leq 1, q > 1$ and concave in $x$ for $q>0, q \neq 1$. 
By the concavity of $\hbox{Ln}_q$, we have
$$
\sum_yp(y) \hbox{Ln}_q(p(x\vert y)) \leq \hbox{Ln}_q\left(\sum_y p(y)p(x\vert y)\right).
$$
Taking the summation of the both sides on the above inequality on $x$, we have
\begin{equation} \label{Dar_eq_01}
\sum_yp(y) \sum_x \hbox{Ln}_q(p(x\vert y)) \leq \sum_x \hbox{Ln}_q(p(x))
\end{equation}
Since $p(y)^q \leq p(y)$ for any $y$ and $q > 1$, and $\hbox{Ln}_q (t) \geq 0$ for any $t \geq 0$ and $q > 1$,
we have,
$$
p(y)^q \sum_x\hbox{Ln}_q(p(x\vert y)) \leq p(y) \sum_x \hbox{Ln}_q(p(x\vert y)).
$$
Taking the summation of the both sides on the above inequality on $y$, we have
\begin{equation} \label{Dar_eq_02}
\sum_yp(y)^q \sum_x \hbox{Ln}_q(p(x\vert y)) \leq \sum_yp(y) \sum_x \hbox{Ln}_q(p(x\vert y)). 
\end{equation}
By Eq.(\ref{Dar_eq_01}) and Eq.(\ref{Dar_eq_02}), we have
$$
\sum_yp(y)^q \sum_x \hbox{Ln}_q(p(x\vert y)) \leq \sum_x \hbox{Ln}_q(p(x)).
$$ 
Therefore we have $ S_q(X\vert Y) \leq S_q(X)$.
The equality holds when $q=1$ and $p(x\vert y) = p(x )$, which means $X$ and $Y$ are independent each other.
Thus the proof of the proposition was completed.

\hfill \qed
\vspace*{2mm}

Eq.(\ref{eq_sub_Dar}) and Eq.(\ref{chain1}) imply the subadditivity of Tsallis entropies.
\begin{The}  {\bf (\cite{Dar})}  \label{subadditivity_theorem}
For $q \geq 1$, we have
\begin{equation} \label{subadditivity}
S_q(X,Y) \leq S_q(X) + S_q(Y).
\end{equation}
\end{The}

From Eq.(\ref{eq_sub_Dar}) and Theorem \ref{chain_entropy}, we have the following extended relation which is often called
{\it independence bound on entropy} in information theory.

\begin{The}
Let $X_1,X_2,\cdots ,X_n$ be the random variables obeying to the probability distribution $p(x_1,x_2,\cdots ,x_n)$.  Then for $q \geq 1$, we have 
\begin{equation} \label{eq_gen_sub_for_mu}
S_q \left( {X_1 , \cdots ,X_n } \right) \le \sum\limits_{i = 1}^n {S_q \left( {X_i } \right)} 
\end{equation}
with equality if and only if $q=1$ the random variables are independent each other.
\end{The}

On the other hand, we easily find that for two independent random variables $X$ and $Y$, and $0 \leq q < 1$, the superadditivity holds:
$$S_q(X\times Y) \geq S_q(X) + S_q(Y).$$
 However, in general the superadditivity for two correlated random variables $X$ and $Y$, and $0 \leq q < 1$ does not hold.
Becasue we can show many counterexamples. For example, we consider the following joint distribution of $X$ and $Y$.
\begin{equation}\label{ex01}
p(x_1,y_1)=p(1-x),p(x_1,y_2)=(1-p)y,p(x_2,y_1)=px,p(x_2,y_2)=(1-p)(1-y),
\end{equation}
where $0 \leq p,x,y \leq 1$.  Then each marginal distribution can be computed by
\begin{equation}\label{ex02}
p(x_1)=p(1-x)+(1-p)y,p(x_2)=px+(1-p)(1-y),p(y_1)=p,p(y_2)=1-p.
\end{equation}
In general, we clearly see $X$ and $Y$ are not independent each other for the above example.
Then the value of $\Delta \equiv S_q(X, Y) - S_q(X) - S_q(Y)$ takes both positive and negative so that there does not exist the complete ordering between
$S_q(X, Y)$  and $S_q(X) + S_q(Y)$ for correlated $X$ and $Y$ in the case of $0 \leq q < 1$. Indeed, $\Delta = -0.287089$ when $q=0.8,p=0.6,x=0.1,y=0.1$, while $\Delta =0.0562961$ when 
$q=0.8,p=0.6,x=0.1,y=0.9$.

Up to the above discussions, we have the possibility that the strong subadditivity holds in the case of $q \geq 1$. 
Indeed we can prove the following strong subadditivity for Tsallis type entropies.
\begin{The}\label{the_ssa}
For $q \geq 1$, the strong subadditivity
\begin{equation} \label{eq_ssa}
S_q \left( {X,Y,Z} \right) + S_q \left( Z \right) \le S_q \left( {X,Z} \right) + S_q \left( {Y,Z} \right)
\end{equation}
holds with equality if and only if $q=1$ and, random variables $X$ and $Y$ are independent for a given random variable $Z$.
\end{The}
{\bf (Proof)}
Theorem is proven by the similar way of Proposition \ref{sub_Dar}.
By the concavity of $\hbox{Ln}_q$, we have
$$
\sum_y p(y\vert z) \hbox{Ln}_q\left(p(x \vert y,z)\right) \leq \hbox{Ln}_q\left(\sum_y p(y\vert z) p(x\vert y,z)\right) 
$$
Multiplying $p(z)^q$ to the both sides of the above inequality and then taking the summation on $z$ and $x$, we have
\begin{equation} \label{eq_ssa01}
\sum_{y,z} p(z)^q p(y\vert z) \sum_x \hbox{Ln}_q(p(x\vert y,z)) \leq \sum_z p(z)^q \sum_x \hbox{Ln}_q(p(x\vert z))
\end{equation}
since $\sum_y p(y\vert z) p(x\vert y,z) = p(x\vert z)$.

For any $y$, $z$ and $q \geq 1$, we have $p(y\vert z)^q \leq p(y\vert z)$. Then by the nonnegativity of the function $\hbox{Ln}_q$, we have
$$
p(y\vert z)^q \sum_x \hbox{Ln}_q(p(x\vert y,z)) \leq p(y\vert z) \sum_x \hbox{Ln}_q(p(x\vert y,z)).
$$
Multiplying $p(z)^q$ to the both sides of the above inequality and then taking the summation on $y$ and $z$, we have
\begin{equation} \label{eq_ssa02}
\sum_{y,z}p(z)^qp(y\vert z)^q \sum_x \hbox{Ln}_q(p(x\vert y,z)) \leq \sum_{y,z}p(z)^q p(y\vert z) \sum_x \hbox{Ln}_q(p(x\vert y,z)).
\end{equation}
From Eq.(\ref{eq_ssa01}) and Eq.(\ref{eq_ssa02}), we have
$$
\sum_{y,z} p(y,z)^q \sum_x \hbox{Ln}_q(p(x\vert y,z)) \leq \sum_z p(z)^q \sum_x \hbox{Ln}_q(p(x\vert z)).
$$
Thus we have 
\begin{equation} \label{con_reduce_ent2}
S_q \left( {X\left| {Y,Z} \right.} \right) \le S_q \left( {X\left| Z \right.} \right),
\end{equation}
which implies
\[
S_q \left( {X,Y,Z} \right) - S_q \left( {Y,Z} \right) \le S_q \left( {X,Z} \right) - S_q \left( Z \right),
\]
by chain rules.
The equality holds when $q=1$ and $p(x\vert y,z) = p(x \vert z)$, which means $X$ and $Y$ are independent each other for a given $Z$.
\hfill \qed
\vspace*{2mm}

We find that Eq.(\ref{eq_ssa}) recovers Eq.(\ref{subadditivity}), by taking the random variable $Z$ as a trivial one. 
This means that Eq.(\ref{eq_ssa}) is a generalization of Eq.(\ref{subadditivity}). We have the further generalized inequality by the similar way of the proof of Theorem \ref{the_ssa}.
\begin{The}
Let $X_1, \cdots ,X_{n+1}$ be the random variables. For $q>1$, we have
\begin{equation} \label{eq_gen_ssa}
S_q(X_{n+1} \vert X_1,\cdots,X_n) \leq S_q(X_{n+1} \vert X_2,\cdots,X_n).
\end{equation} 
\end{The}

The subadditivity for Tsallis entropies conditioned by $Z$ holds.
\begin{Prop} \label{prop_sa_cond}
For $q \geq 1$, we have
\begin{equation}
S_q(X,Y\vert Z) \leq S_q(X\vert Z) + S_q(Y\vert Z).
\end{equation}
\end{Prop}
{\bf (Proof)}
Summing $-2 S_q(Z)$ to the both sides in Eq.(\ref{eq_ssa}), we have
$$
S_q(X,Y,Z) - S_q(Z) \leq S_q(X,Z) - S_q(Z) + S_q(Y,Z) -S_q(Z).
$$
By chain rules, we have the proposition.
\hfill \qed
\vspace*{2mm}

Proposition \ref{prop_sa_cond} can be generalized in the following.

\begin{The}
For $q \geq 1$, we have
\begin{equation}
S_q(X_1,\cdots,X_n\vert Z) \leq S_q(X_1\vert Z) + \cdots + S_q(X_n\vert Z).
\end{equation}
\end{The}

In addition, we have the following propositions.
\begin{Prop}
For $q \geq 1$, we have
$$2 S_q(X,Y,Z) \leq S_q(X,Y) + S_q(Y,Z) + S_q(Z,X).$$
\end{Prop}
{\bf (Proof)}
Form Eq.(\ref{eq_sub_Dar}) and Eq.(\ref{con_reduce_ent2}), we have
$$S_q(X\vert Y,Z) \leq S_q(X), $$
and 
$$S_q(Z\vert X,Y) \leq S_q(Z\vert X). $$
Thus we have
$$ S_q(X\vert Y,Z)  +S_q(Z\vert X,Y) \leq S_q(X) + S_q(Z\vert X).  $$
By chain rules, we have
$$ S_q(X,Y,Z) - S_q(X,Y) + S_q(X,Y,Z) - S_q(Y,Z) \leq S_q(Z,X), $$
which implies the proposition.
\hfill \qed
\vspace*{2mm}

\begin{Prop} \label{prop_dist01}
For $q >1$, we have
$$
S_q(X_n\vert X_1) \leq S_q(X_2 \vert X_1) + \cdots + S_q(X_n \vert X_{n-1}).
$$
\end{Prop}
{\bf (Proof)}
From Eq.(\ref{chain_gen}) and Eq.(\ref{con_reduce_ent2}), we have
$$
S_q(X_1,\cdots,X_n) \leq S_q(X_1) + S_q(X_2\vert X_1) + \cdots S_q(X_n\vert X_{n-1}).
$$
Therefore we have
\begin{eqnarray*}
S_q(X_n\vert X_1) &\leq& S_q(X_2,\cdots,X_n\vert X_1) \\
&=& S_q(X_1,\cdots,X_n) -S_q(X_1) \\
&\leq & S_q(X_2\vert X_1) + \cdots +S_q(X_n\vert X_{n-1}),
\end{eqnarray*}
by Eq.(\ref{chain1}) and Eq.(\ref{eq_lem22_result}).
\hfill \qed
\vspace*{2mm}


\section{Tsallis mutual entropy}   \label{sec4}
For normalized Tsallis entropies, the mutual information was defined in \cite{Ya} with the assumption of its nonegativity. 
We define the Tsallis mutual entropy in terms of the original ({\it not} normalized) Tsallis type entropies.
The inequality Eq.(\ref{eq_sub_Dar}) naturally leads us to define Tsallis mutual entropy without the assumption of its nonegativity.

\begin{Def}
For two random variables $X$ and $Y$, and $q > 1$, we define the Tsallis mutual entropy 
as the difference between Tsallis entropy and Tsallis conditional entropy such that
\begin{equation} \label{def_mutual_Tsallis}
I_q(X;Y) \equiv S_q(X) - S_q(X\vert Y).
\end{equation}
\end{Def}
Note that we never use the term {\it mutual information} but use mutual entropy through this paper,
 since a proper evidence of channel coding theorem for information transmission has not ever been shown in the context of Tsallis
statistics.
From Eq.(\ref{chain1}), Eq.(\ref{subadditivity}) and Eq.(\ref{eq_sub_Dar}), we easily find that $I_q(X;Y)$ has the following fundamental properties.

\begin{Prop}
\begin{itemize}
\item[(1)] $0 \leq I_q(X;Y) \leq min \left\{S_q(X), S_q(Y)\right\}.$
\item[(2)] $ I_q(X;Y) = S_q(X)+ S_q(Y) - S_q(X,Y) = I_q(Y;X).$
\end{itemize}
\end{Prop}

Note that we have
\begin{equation} \label{ordering_tme}
S_q(X) \leq S_q(Y) \Longleftrightarrow S_q(X\vert Y) \leq S_q(Y\vert X)
\end{equation}
from the symmetry of Tsallis mutual entropy.
We also define the Tsallis conditional mutual entropy
\begin{equation}
I_q(X;Y\vert Z) \equiv S_q(X\vert Z) - S_q(X\vert Y,Z)
\end{equation}
for three random variables $X$, $Y$ and $Z$, and $q > 1$.

In addition, $I_q(X;Y\vert Z)$ is nonnegative, thanks to Eq.(\ref{con_reduce_ent2}). 
For these quantities, we have the following chain rules.

\begin{The}
\begin{itemize}
\item[(1)] For three random variables $X$, $Y$ and $Z$, and $q > 1$, the chain rule holds:
\begin{equation}\label{chain_mutual1}
I_q(X;Y,Z) = I_q(X;Z) + I_q(X;Y\vert Z).
\end{equation}
\item[(2)] For random variables $X_1,\cdots ,X_n$ and $Y$, the chain rule holds:
\begin{equation}
I_q(X_1,\cdots ,X_n;Y) = \sum_{i=1}^n I_q(X_i; Y\vert X_1,\cdots ,X_{i-1}).
\end{equation}
\end{itemize}
\end{The}

{\bf (Proof)}
(1) follows from the chain rules of Tsallis entropy.
\begin{eqnarray*}
 I_q \left( {X;Y\left| Z \right.} \right) &=& S_q \left( {X\left| Z \right.} \right) - S_q \left( {X\left| {Y,Z} \right.} \right) \\ 
&=& S_q \left( {X\left| Z \right.} \right) - S_q \left( X \right) + S_q \left( X \right) - S_q \left( {X\left| {Y,Z} \right.} \right) \\ 
&=&  - I_q \left( {X;Z} \right) + I_q \left( {X;Y,Z} \right). 
\end{eqnarray*}

(2) follows from the application of Eq.(\ref{chain_gen}) and Eq.(\ref{eq_remark23}).
\begin{eqnarray*}
 I_q \left( {X_1 , \cdots ,X_n ;Y} \right) &=& S_q \left( {X_1 , \cdots ,X_n } \right) - S_q \left( {X_1 , \cdots ,X_n \left| Y \right.} \right) \\ 
 &=& \sum\limits_{i = 1}^n {S_q \left( {X_i \left| {X_{i - 1} , \cdots ,X_1 } \right.} \right)}  - \sum\limits_{i = 1}^n {S_q \left( {X_i \left| {X_{i - 1} , \cdots ,X_1 ,Y} \right.} \right)}  \\ 
 &=& \sum\limits_{i = 1}^n {I_q \left( {X_i ;\left. Y \right|X_1 , \cdots ,X_{i - 1} } \right)} . 
\end{eqnarray*}

\hfill \qed
\vspace*{2mm}

We have the following inequality for Tsallis mutual entropies by the strong subadditivity.
\begin{Prop}
For $q > 1$, we have $$I_q(X;Z) \leq I_q(X,Y;Z).$$
\end{Prop}

\section{Correlation coefficients and distances for Tsallis entropies}    \label{sec5}

We discuss on the entropic distance between $X$ and $Y$ by means of Tsallis conditional entropy $S_q(X\vert Y)$.
If $S_q(X\vert Y) = 0 = S_q(Y\vert X)$, then $X$ and $Y$ are called {\it equivalence} and denoted by $X \equiv Y$. 
Defining an entropic distance between $X$ and $Y$ as 
\begin{equation}
d_q(X,Y) \equiv S_q(X\vert Y) + S_q(Y\vert X),
\end{equation}
we have the following proposition.

\begin{Prop}
For $q > 1 $, $d_q$ satisfies
\begin{itemize}
\item[(i)]   $d_q(X,Y) = 0$ if and only if $X \equiv Y$.
\item[(ii)] $d_q(X,Y) = d_q(Y,X)$.
\item[(iii)] $d_q(X,Z) \leq d_q(X,Y) + d_q(Y,Z)$.
\end{itemize}
\end{Prop}
{\bf (Proof)}
(i) and (ii) are clear from the definition of $d_q$.
In the case of $n=3$ in Proposition \ref{prop_dist01}, we have the triangle inequality:
$$
S_q(X_3\vert X_1) \leq S_q(X_2 \vert X_1) + S_q(X_3 \vert X_{2}).
$$
Putting $X_1 = Z$,$X_2=Y$ and $X_3=X$ in the aboce inequality, we have
\begin{equation} \label{dist_eq_02}
S_q(X\vert Z) \leq S_q(Y \vert Z) + S_q(X \vert Y).
\end{equation}
In addition, exchanging $X$ and $Z$ in the above inequality, we have
$$
S_q(Z\vert X) \leq S_q(Y \vert X) + S_q(Z \vert Y).
$$
Summing the both sides of two above inequalities, we have the proposition.
\hfill \qed
\vspace*{2mm}

$I_q(X;Y)$ represents a kind of correlation between $X$ and $Y$. Following \cite{Raj}, 
we define a parametrically extended correlation coefficient in terms of Tsallis mutual entropy
such that
\begin{equation}
\rho_q(X,Y) \equiv \frac{I_q(X;Y)}{S_q(X,Y)}
\end{equation}
for $S_q(X) > 0$, $S_q(Y) > 0$ and $q >1$.
Then we have the following proposition.

\begin{Prop}  \label{4sec_ent_dist_prop2}
For $q > 1$, $S_q(X) > 0$ and $S_q(Y) >0$, we have the following properties.
\begin{itemize}
\item[(i)] $\rho_q(X,Y) = \rho_q(Y,X)$.
\item[(ii)] $0 \leq \rho_q(X,Y) \leq 1$.  
\item[(iii)] $\rho_q(X,Y) =0 $ if and only if $X$ and $Y$ are independent each other and $q=1$.
\item[(iv)] $\rho_q(X,Y) = 1$ if and only if $X \equiv Y$.
\end{itemize}
\end{Prop}
{\bf (Proof)}
(i) is clear from the definition of $\rho_q$.
(ii) follows from $0 \leq I_q(X;Y) \leq S_q(X) \leq S_q(X,Y)$.
(iii) follows from Proposition \ref{sub_Dar}. (iv) follows from $I_q(X;Y) = S_q(X,Y) \Longleftrightarrow S_q(X\vert Y) = S_q(Y\vert X) = 0$.

\hfill \qed
\vspace*{2mm}

Again, we define an entropic distance between $X$ and $Y$ by
\begin{equation}
\widetilde d_q(X,Y) \equiv 1- \rho_q(X,Y).
\end{equation}
Then we have the following proposition.

\begin{Prop}
For $q > 1$, $S_q(X) > 0$, $S_q(Y) >0$ and $S_q(Z) > 0$, we have the following properties.
\begin{itemize}
\item[(i)] $\widetilde d_q(X,Y) = 0 $ if and only if $X \equiv Y$.
\item[(ii)] $\widetilde d_q(X,Y) = \widetilde d_q(Y,X).$
\item[(iii)] $\widetilde d_q(X,Z) \leq \widetilde d_q(X,Y) + \widetilde d_q(Y,Z)$.
\end{itemize}
\end{Prop}
{\bf (Proof)}
(i) and (ii) follows from (iv) and (i) of Proposition \ref{4sec_ent_dist_prop2}, respectively.
From chain rules and Eq.(\ref{dist_eq_02}), we have
\begin{eqnarray}
\frac{S_q(X\vert Z)}{S_q(X,Z)} &=& \frac{S_q(X\vert Z)}{S_q(X\vert Z) + S_q(Z)} \nonumber \\
& \leq & \frac{S_q(X\vert Y) + S_q(Y\vert Z) }{S_q(X\vert Y) + S_q(Y\vert Z) + S_q(Z)} \nonumber  \\
&=& \frac{S_q(X\vert Y) }{S_q(X\vert Y) + S_q(Y\vert Z) + S_q(Z)}   + \frac{S_q(Y\vert Z) }{S_q(X\vert Y) + S_q(Y\vert Z) + S_q(Z)}  \nonumber  \\
& \leq & \frac{S_q(X\vert Y) }{S_q(X\vert Y) + S_q(Z)}   + \frac{S_q(Y\vert Z) }{ S_q(Y\vert Z) + S_q(Z)}  \nonumber  \\
& =& \frac{S_q(X\vert Y)}{S_q(X,Y)} +\frac{S_q(Y\vert Z)}{S_q(Y,Z)}.  \label{eq_dist_03}
\end{eqnarray}
Exchanging $X$ and $Z$, we have 
\begin{equation}
\frac{S_q(Z\vert X)}{S_q(Z,X)} \leq \frac{S_q(Z\vert Y)}{S_q(Z,Y)} +\frac{S_q(Y\vert X)}{S_q(Y,X)}.  \label{eq_dist_04}
\end{equation}
Summing the both sides of two inequalities Eq.(\ref{eq_dist_03}) and Eq.(\ref{eq_dist_04}), we have (iii), since Tsallis joint entropy is symmetry.

\hfill \qed
\vspace*{2mm}

We also define another correlation coefficient in terms of Tsallis mutual entropy by
\begin{equation} 
\hat \rho_q(X,Y) \equiv \frac{I_q(X;Y)}{max\left\{S_q(X),S_q(Y)\right\}}
\end{equation}
for $S_q(X) > 0$, $S_q(Y) > 0$ and $q >1$. See \cite{Horibe}.
Then we have the following proposition. 

\begin{Prop}\label{4sec_ent_dist_prop3}
For $q > 1$, $S_q(X) > 0$ and $S_q(Y) >0$, we have the following properties.
\begin{itemize}
\item[(i)] $\hat \rho_q(X,Y) = \hat \rho_q(Y,X)$.
\item[(ii)] $0 \leq \hat \rho_q(X,Y) \leq 1$.  
\item[(iii)] $\hat \rho_q(X,Y) =0 $ if and only if $X$ and $Y$ are independent each other and $q=1$.
\item[(iv)] $\hat \rho_q(X,Y) = 1$ if and only if $X \equiv Y$.
\end{itemize}
\end{Prop}
{\bf (Proof)}
(i)-(iv) follow straightforwardly.

\hfill \qed
\vspace*{2mm}

We also define an entropic distance between $X$ and $Y$ by
\begin{equation}
\hat d_q(X,Y) \equiv 1- \hat \rho_q(X,Y).
\end{equation}
Then we have the following proposition.

\begin{Prop}
For $q > 1$, $S_q(X) > 0$, $S_q(Y) >0$ and $S_q(Z) > 0$, we have the following properties.
\begin{itemize}
\item[(i)] $\hat d_q(X,Y) = 0 $ if and only if $X \equiv Y$.
\item[(ii)] $\hat d_q(X,Y) = \hat d_q(Y,X).$
\item[(iii)] $\hat d_q(X,Z) \leq \hat d_q(X,Y) + \hat d_q(Y,Z)$.
\end{itemize}
\end{Prop}
{\bf (Proof)}
(i) and (ii) follows from Proposition \ref{4sec_ent_dist_prop3}.
(iii) is proven by the similar way in \cite{Horibe}.
In order to show (iii), we can assume $S_q(Z) \leq S_q(X)$ without loss of generality.
Then we prove (iii) in the following three cases.
\begin{itemize}
\item[(a)] $S_q(Y) \leq S_q(Z) \leq S_q(X)$ :
From Eq.(\ref{ordering_tme}) and Eq.(\ref{dist_eq_02}), we have
\begin{eqnarray*}
\hat d_q(X,Z) = \frac{S_q(X\vert Z)}{S_q(X)} &\leq& \frac{S_q(X\vert Y)}{S_q(X)} + \frac{S_q(Y\vert Z)}{S_q(X)} 
\leq \frac{S_q(X\vert Y)}{S_q(X)} + \frac{S_q(Y\vert Z)}{S_q(Z)} \\
&\leq & \frac{S_q(X\vert Y)}{S_q(X)} + \frac{S_q(Z\vert Y)}{S_q(Z)}  = \hat d_q(X,Y) + \hat d_q(Y,Z).
\end{eqnarray*}
\item[(b)]  $S_q(Z) \leq S_q(X) \leq S_q(Y)$ :
It is shown as similar as (a).
\item[(c)]  $S_q(Y) \leq S_q(Z) \leq S_q(X)$ :
From  Eq.(\ref{dist_eq_02}), we have
\begin{eqnarray*}
\hat d_q(X,Z) &=& \frac{S_q(X\vert Z)}{S_q(X)} = 1-  \left(1- \frac{S_q(X\vert Z)}{S_q(X)}  \right) \leq 1- \frac{S_q(X)}{S_q(Y)}\left(1- \frac{S_q(X\vert Z)}{S_q(X)}  \right) \\
&=& \frac{S_q(Y)-S_q(X)+S_q(X\vert Z)}{S_q(Y)} \leq \frac{S_q(Y)-S_q(X)+S_q(X\vert Y)+S_q(Y\vert Z)}{S_q(Y)} \\
&=& \hat d_q(X,Y) + \hat d_q(Y,Z).
\end{eqnarray*}
\end{itemize}

\hfill \qed
\vspace*{2mm}

\section{Applications of Tsallis entropies}   \label{sec6}
\subsection{Generalized Fano's inequality}
As an application of Tsallis conditional entropy, we give a generalized Fano's inequality.
\begin{The}
For $q > 1$ and two random variables taking values in the finite (alphabet) set $\fX$, the following inequality holds.
$$
S_q(X\vert Y) \leq P(X\neq Y)\ln_q(\vert \fX\vert -1) +s_q(P(X\neq Y)),
$$
where $s_q$ represents the Tsallis binary entropy:
$$
s_q(p) \equiv -p^q\ln_q p -(1-p)^q \ln_q (1-p).
$$
\end{The}
{\bf (Proof)}
Define the random variables $Z$ by
\[
Z = \left\{ \begin{array}{l}
 0\,\,\,\,\,\,\,\,\,\,\,\,\,\,\,\,{\rm if}\,\,\,\,X = Y \\ 
 {\rm 1}\,\,\,\,\,\,\,\,\,\,\,\,\,\,\,\,\,{\rm if}\,\,\,\,X \ne Y .\\ 
 \end{array} \right.\,\,\,
\]
By chain rules, $S_q(Z\vert X,Y) = 0$ and $S_q(Z\vert Y) \leq S_q(Z) =s_q(P(X\neq Y))$, we then have 
\begin{eqnarray*}
S_q(X\vert Y) &=& S_q(X\vert Y) + S_q(Z\vert X,Y)\\
&=&S_q(X,Z\vert Y)\\
&=& S_q(X\vert Y,Z) + S_q(Z\vert Y) \\
&\leq & S_q(X\vert Y,Z) + S_q(Z) \\
&=& S_q(X\vert Y=y,Z=0) + S_q(X\vert Y=y,Z=1)+ s_q(P(X\neq Y)) \\
&=& S_q(X\vert Y=y,Z=1)+ s_q(P(X\neq Y)) \\
& \leq & \ln_q(X\neq Y)+ s_q(P(X\neq Y)),
\end{eqnarray*}
since the nonnegativity of Tsallis relative entropy implies $ -\sum_{i=1}^n p_i^q \ln_qp_i \leq \ln_q n$.

\hfill \qed
\vspace*{2mm}

\subsection{Entropy rate for Tsallis entropy}
Moreover, we discuss on the Tsallis entropy rate.

\begin{Def}
The Tsallis entropy rate of a stochastic process ${\bf X} \equiv \left\{ X_i \right\}$ is defined by
\[
S_q \left( {\bf X} \right) \equiv \mathop {\lim }\limits_{n \to \infty } \frac{{S_q \left( {X_1 , \cdots ,X_n } \right)}}{n}
\]
if the limit exists.
\end{Def}

$S_q \left( {\bf X} \right)$ can be considered as Tsallis entropy per a symbol in the limit.

\begin{The}
For $q > 1$ and a stationary stochastic process, the Tsallis entropy rate exsits and is expressed by
\[
S_q \left( {\bf X} \right) = \mathop {\lim }\limits_{n \to \infty } S_q \left( {X_n \left| {X_1 , \cdots ,X_{n - 1} } \right.} \right).
\]
\end{The}

{\bf (Proof)}

By the stationary process and the inequality Eq.(\ref{eq_gen_ssa}), we have
\begin{eqnarray*}
 S_q \left( {X_n \left| {X_1 , \cdots ,X_{n - 1} } \right.} \right) &=& S_q \left( {X_{n + 1} \left| {X_2 , \cdots ,X_n } \right.} \right) \\ 
 &\ge &S_q \left( {X_{n + 1} \left| {X_1 ,X_2 , \cdots ,X_n } \right.} \right). 
\end{eqnarray*}

Thus $ S_q \left( {X_n \left| {X_1 , \cdots ,X_{n - 1} } \right.} \right) $ converges to a certain value $s$, 
since it is a monotone decreasing sequence with respect to $n$.
That is,
$$
\mathop {\lim }\limits_{n \to \infty } S_q \left( {X_n \left| {X_1 , \cdots ,X_{n - 1} } \right.} \right) = s.
$$
Moreover by the chain rule and Ces\'aro's theorem (e.g., see pp.64 of \cite{Cov}), we have
\[
S_q({\bf X}) \equiv \mathop {\lim }\limits_{n \to \infty } \frac{{S_q \left( {X_1 , \cdots ,X_n } \right)}}{n} = \mathop {\lim }\limits_{n \to \infty } \sum\limits_{i = 1}^n {\frac{{S_q \left( {X_i \left| {X_1 , \cdots ,X_{i - 1} } \right.} \right)}}{n}}  = s.
\]

\hfill \qed
\vspace*{2mm}

\section{Remarks and discussions} \label{sec7}
\subsection{Chain rule and subadditivity for nonadditive entropies}
In \cite{Su3}, the nonadditive entropies including Tsallis entropy and the entropy of type $\beta$ was axiomatically characterized by
the generalized Shannon-Khinchin's axiom in the following manner.
The function $\widetilde S_q(x_1, \cdots ,x_n)$ is assumed to be defined for the $n$-tuple $(x_1, \cdots ,x_n)$ belonging to 
$$\Delta _n \equiv \left\{ (p_1,\cdots ,p_n) \vert \sum_{i=1}^n p_i =1, p_i \geq 0\,\, (i=1,2,\cdots ,n)\right\}$$ 
and to take a value in $\hbox{R}^+ \equiv [0,\infty )$.
Then it was shown in \cite{Su3} that four conditions {\it continuity}, {\it maximality}, {\it generalized Shannon additivity} and {\it expandability} in \cite{Su3} determine the function $\widetilde S_q : \Delta_n \to \hbox{R}^+$ such that
\begin{equation} \label{this_generalization}
\widetilde S_q(x_1, \cdots ,x_n) = \frac{\sum_{i=1}^n \left(x_i^q -x_i\right)}{\phi(q)},
\end{equation}
where $\phi(q)$ is characterized by the following conditions:
\begin{itemize}
\item[(i)] $\phi(q)$ is continuous and $\phi(q) (1-q) >0$ for $q \neq 1$.
\item[(ii)] $\lim_{q \to 1} \phi(q) = 0 $ and $\phi(q) \neq 0$ for $q \neq 1$.
\item[(iii)] There exists $(a,b) \subset \hbox{R}^+ $ such that $a < 1 <b$ and $\phi(q)$
is differentiable on $(a,1)$ and $(1,b)$. 
\item[(iv)] There exists a positive constant number $k$ such that $\lim_{q \to 1}\frac{d\phi(q)}{dq} = -\frac{1}{k}$.
\end{itemize}
Simple calculations show the nonextensivity
$$
\widetilde S_q(X \times Y) = \widetilde S_q(X)  + \widetilde S_q(Y)  + \phi(q) \widetilde S_q(X)   \widetilde S_q(Y)  
$$
for the nonadditive entropy $\widetilde S_q$.
From the view of the generalization Eq.(\ref{this_generalization}), we have the following chain rule.
\begin{Lem} \label{nonex_chain}
For the nonadditive entropy 
$$\widetilde S_q(X) \equiv  \frac{\sum_{x} \left(p(x)^q -p(x)\right)}{\phi(q)}$$
and any continuous function $\phi(q)$, we have the chain rule
$$
\widetilde S_q(X,Y) =\widetilde S_q(X) + \widetilde S_q(Y\vert X), 
$$ 
where the nonadditive joint entropy and the nonadditive conditional entropy are defined by
$$
\widetilde S_q(X,Y) \equiv \frac{\sum_{x,y} \left(p(x,y)^q -p(x,y)\right)}{\phi(q)}
$$
and 
$$
\widetilde S_q(X\vert Y) \equiv \sum_y p(y)^q \widetilde S_q(X\vert y) = \sum_y p(y)^q \left\{\frac{\sum_{x} \left(p(x\vert y)^q -p(x\vert y)\right)}{\phi(q)} \right\},
$$
respectively.
\end{Lem}
{\bf (Proof)}
It follows immediately.
\begin{eqnarray*}
\widetilde S_q(Y\vert X) &=& \frac{\sum_{y,x}p(x)^qp(y\vert x)^q-\sum_xp(x)^q}{\phi(q)}\\
&=&\frac{\sum_{x,y}\left(p(x,y)^q-p(x,y)\right)}{\phi(q)} -\frac{\sum_x \left(p(x)^q-p(x)\right)}{\phi(q)}\\
&=& \widetilde S_q(X,Y) - \widetilde S_q(X).
\end{eqnarray*}
\hfill \qed
\vspace*{2mm}

From the view of the generalization Eq.(\ref{this_generalization}), Proposition \ref{sub_Dar} also generalized as follows.
\begin{Lem}  \label{nonex_reduced}
For the nonadditive entropy $\widetilde S_q(X) $
where $\phi(q)$ is continuous and $\phi(q) (1-q) >0$ for $q \neq 1$,
we have the inequality :
$$\widetilde S_q(X\vert Y) \leq \widetilde S_q(X).$$
\end{Lem}

{\bf (Proof)}
Since the function $\widetilde{\hbox{Ln}}_q(x) \equiv \frac{x^q-x}{\phi(q)}, (q \geq 0,q \neq 1, x \geq 0)$ is 
nonnegative for $0 \leq x \leq 1, q > 1$ and concave in $x$ for $q >0, q \neq 1$, the theorem follows as similar as 
the proof of Proposition \ref{sub_Dar}.

\hfill \qed
\vspace*{2mm}
 
Form Lemma \ref{nonex_chain} and Lemma \ref{nonex_reduced}, we have the subadditivity for nonadditive entropy.

\begin{The} {\bf (Subadditivity for nonadditive entropy)}
For the nonadditive entropy
$\widetilde S_q(X)$ where $\phi(q)$ is continuous and $\phi(q) (1-q) >0$ for $q \neq 1$,
we have the subadditivity :
$$\widetilde S_q(X,Y) \leq \widetilde S_q(X) +  \widetilde S_q(Y) .$$
\end{The}

We note that we need the condition that $\phi(q) (1-q) >0$ for $q \neq 1$ to prove Lemma \ref{nonex_reduced}, 
while we do not need any condition to prove Lemma \ref{nonex_chain}.

\subsection{Inequalities on pseudoadditivity}

The pseudoadditivity Eq.(\ref{pseudo}) for independent random variables $X$ and $Y$ gives rise to the mathematical
interest whether
we have the complete ordering between the left hand side
and the right hand side
in Eq.(\ref{pseudo}) for two general random variables $X$ and $Y$. Such a kind of
problem was taken in the paper 
\cite{Fa} for the normalized Tsallis type entropies which are
different from 
the definitions of the Tsallis type entropies in the present paper.
However, its inequality appeared in (3.5) of the paper \cite{Fa}
 was not true as we found the counter example in \cite{SF}. 

Unfortunately, in the present case, we also find the
counter example for the inequalities between $S_q(X,Y)$ and $S_q(X) + S_q(Y)
+(1-q) S_q(X)S_q(Y)$.
In the same setting of Eq.(\ref{ex01}) and Eq.(\ref{ex02}), 
$\delta \equiv S_q(X,Y) - \left\{S_q(X) + S_q(Y) +(1-q)
S_q(X)S_q(Y) \right\}$ takes
both positive and negative values for both cases
$0\leq q<1$ and $q >1$.
Indeed, $\delta = 0.00846651$ when $q=1.8,p=0.1,x=0.1,y=0.8$, while
$\delta = -0.0118812$ when $q=1.8,p=0.1,x=0.8,y=0.1$.
Also, $\delta = 0.00399069$ when $q=0.8,p=0.1,x=0.8,y=0.1$, while
$\delta = -0.0128179$ when $q=0.8,p=0.1,x=0.1,y=0.8$.

Therefore there does not exist the complete ordering 
between $S_q(X,Y)$ and $S_q(X) + S_q(Y) +(1-q)
S_q(X)S_q(Y)$ for both cases $0 \leq q < 1$ and $ q > 1 $.

\subsection{A remarkable inequality derived from subadditivity for Tsallis entropies}
From Eq.(\ref{subadditivity}), we have the following inequality
\begin{equation} \label{change_subadditivity_of_Tsallis}
\sum\limits_{i = 1}^n {\left( {\sum\limits_{j = 1}^m {p_{ij} } } \right)} ^r  + \sum\limits_{j = 1}^m {\left( {\sum\limits_{i = 1}^n {p_{ij} } } \right)} ^r 
 \le \sum\limits_{i = 1}^n {\sum\limits_{j = 1}^m {p_{ij} ^r } }  + \left( {\sum\limits_{i = 1}^n {\sum\limits_{j = 1}^m {p_{ij} } } } \right)^r 
\end{equation}
for $r \geq 1 $ and $p_{ij}$  satisfying $0 \leq p_{ij} \leq 1$ and  $\sum_{i=1}^n\sum_{j=1}^m p_{ij} =1$.
By putting $p_{ij} = \frac{a_{ij}}{\sum_{i=1}^n\sum_{j=1}^ma_{ij}}$ in Eq.(\ref{change_subadditivity_of_Tsallis}), we have
the following inequality as a corollary of Theorem \ref{subadditivity_theorem}.
\begin{Cor} 
For $r \geq 1 $ and $a_{ij} \geq 0$,
\begin{equation}
\sum\limits_{i = 1}^n {\left( {\sum\limits_{j = 1}^m {a_{ij} } } \right)} ^r  + \sum\limits_{j = 1}^m {\left( {\sum\limits_{i = 1}^n {a_{ij} } } \right)} ^r  \le \sum\limits_{i = 1}^n {\sum\limits_{j = 1}^m {a_{ij} ^r } }  + \left( {\sum\limits_{i = 1}^n {\sum\limits_{j = 1}^m {a_{ij} } } } \right)^r. 
\end{equation}
\end{Cor}
It is remarkable that the following inequality holds \cite{HLP}
\begin{equation} 
\sum\limits_{i = 1}^n {\left( {\sum\limits_{j = 1}^m {a_{ij} } } \right)} ^r  \ge \sum\limits_{i = 1}^n {\sum\limits_{j = 1}^m {a_{ij} ^r } } 
\end{equation}
for $r \geq 1 $ and $a_{ij} \geq 0$.

\subsection{Difference between Tsallis entropy and Shannon entropy}

We point out on the difference between Tsallis entropy and Shannon entropy from the viewpoint of mutual entropy. 
In the case of $q=1$, the relative entropy between the joint probability $p(x,y)$ and the direct probability $p(x)p(y)$ is equal to the mutual entropy:
$$ D_1((X,Y) \vert X\times Y) = S_1(X) - S_1(X\vert Y).$$
However, in the general case ($q \neq 1$), there exists the following relation:
\begin{eqnarray} 
&&D_q((X,Y) \vert X\times Y) = S_q(X) - S_q(X\vert Y) \nonumber \\
&& + \sum_{i,j} p(x_i,y_j) \left\{   p(x_i)^{q-1}\ln_qp(x_i)+ p(y_j)^{q-1}\ln_qp(y_j) -p(x_i,y_j)^{q-1}\ln_q p(x_i)p(y_j) \right\}, \label{third}
\end{eqnarray}
which gives the crucial difference between the special case ($q = 1$) and the general case ($q \neq 1$).
The third term of the right hand side in the above equation Eq.(\ref{third}) vanishes if $q=1$.
The existence of the third term of Eq.(\ref{third}) means that we have two possibilities of the definition of Tsallis mutual entropy, that is, 
$I_q(X;Y) \equiv S_q(X) -S_q(X\vert Y)$ or $I_q(X;Y) \equiv D_q((X,Y) \vert X\times Y)$.
We have adopted the former definition in the present paper, along with the definition of the capacity in the origin of information theory by Shannon \cite{Shannon}.

\subsection{Another candidate of Tsallis conditional entropy }
It is remarkable that Tsallis entropy $S_q(X)$ can be regarded as the expected value of $\ln_q \frac{1}{p(x)}$, that is, since $\ln_q(x) = -x^{1-q} \ln_q (1/x)$, 
it is expressed by
\begin{equation}\label{view}
S_q(X) = \sum_{x} p(x) \ln_q \frac{1}{p(x)},\quad (q\neq 1),
\end{equation}
where the convention $0 \ln_q(\cdot) =0$ is set.
Along with the view of Eq.(\ref{view}), we may define Tsallis conditional entropy and Tsallis joint entropy in the following.
\begin{Def}
For the conditional probability $p(x\vert y)$ and the joint probability $p(x,y)$, we define Tsallis conditional entropy and Tsallis joint entropy by
\begin{equation}  
\hat S_q(X\vert Y) \equiv  \sum_{x,y} p(x,y) \ln_q \frac{1}{p(x\vert y)},  \quad (q \neq 1), \label{new_jc} 
\end{equation}
and
\begin{equation}
 S_q(X,Y) \equiv  \sum_{x,y} p(x,y) \ln_q \frac{1}{p(x,y)}, \quad (q \neq 1).  \label{new_jc2} 
\end{equation}
\end{Def}
We should note that Tsallis conditional entropy defined in Eq.(\ref{new_jc}) is not equal to that defined in Eq.(\ref{jc}), 
while Tsallis joint entropy defined in Eq.(\ref{new_jc2}) is equal to that defined in Eq.(\ref{jc2}). 
If we adopt the above definitions  Eq.(\ref{new_jc}) instead of Eq.(\ref{jc}), we have the following inequality.
\begin{Prop}
For $q>1$, we have 
$$ S_q(X,Y) \leq S_q(X) + \hat S_q(Y\vert X). $$
For $0 \leq q <1$, we have
$$ S_q(X,Y) \geq S_q(X) + \hat S_q(Y\vert X). $$
\end{Prop} 
Therefore we do not have the chain rule for $\hat S_q(Y\vert X)$ in general, 
namely we are not able to construct a parametrically extended information theory in this case.

\section{Conclusion}

As we have seen, we have found that the chain rules for the Tsallis type entropies hold as similar way of the proofs of the entropy of type $\beta$.
This result was generalized from a few insights.
Also we have proved the strong subadditivity of the Tsallis type entropies for the general two random variables in the case of $q > 1$.
Moreover it has been shown that in general the superadditivity of Tsallis entropies  does not hold in the case of $0 \leq q < 1$.
Thus we could give important results for Tsallis entropies in the case of $q > 1$ from the information theoretical point of view. 
Therefore we have the possibility that the parametrically extended information theory will be constructed by starting from
Tsallis entropy for $q > 1$. In other words, there is less possibility of such a construction in the case of $0 \leq q <1$ if our stage still 
in the usual probability theory, because the relation which should be satisfied in the contrast to the usual information theory does not hold in the case of $0 \leq q < 1$. 
This gives us the rough meaning of the parameter $q$. 
That is, we found the crucial different results between $S_q(X)$ for $q > 1$ and $S_q(X)$ for $0 \leq q < 1$ from the information theoretical point of view. 
For $q > 1$, we showed that the Tsallis type entropies have the similar results to Shannon entropies. In other words, it has been shown that
 the definition of Tsallis conditional entropy Eq.(\ref{jc}) and Tsallis mutual entropy Eq.(\ref{def_mutual_Tsallis}) 
have a possibility to be a pioneer to construct a nonadditive information theory.

Our results in the present paper are fundamental so that we are convinced that 
these results will help the progresses of nonextensive statistical physics and information theory. 
We are now studying the coding problem in nonadditive information theory and searching for the precise meaning of the parameter $q$.


\section*{Acknowledgement}
The authour would like to thank Professor K.Yanagi and Professor K.Kuriyama for providing valuable comments.
We also would like to thank the reviewers for providing valuable comments to improve the manuscript.
This work was supported by the Japanese Ministry of Education, Science, Sports and Culture, Grant-in-Aid for 
Encouragement of Young scientists (B), 17740068.

\end{document}